\documentstyle[preprint,eqsecnum,epsfig,prd,aps,floats]{revtex}
\begin{document}

\newcommand{\nl}{\nonumber \\}
\newcommand{\eq}[1]{Eq.~(\ref{#1})}
\newcommand{\npol}{\bigtriangleup}
\newcommand{\dpol}{\bigtriangledown}
\newcommand{\cstar}{{\bf C}^\ast}

\preprint{\tighten{ \vbox{\hbox{PUPT-1862}
		\hbox{MIT-CTP-2867}
                \hbox{hep-th/9906020}  }}}
 
\title{Mirror symmetry for ${\cal N}=1$ QED in three dimensions}
 
\author{ Martin Gremm\thanks{email: gremm@feynman.princeton.edu }\thanks{ 
On leave of absence from MIT, Cambridge, MA 02139}} 
  
\address{Joseph Henry Laboratories, Princeton University, Princeton, NJ 08544} 

\author{Emanuel Katz\thanks{amikatz@mit.edu}}

\address{Center for Theoretical Physics,\\
Massachusetts Institute of Technology, Cambridge, MA 02139}

\maketitle

\begin{abstract}
We construct three-dimensional ${\cal N}=1$ QED with $N_f$ flavors using branes
of type IIB string theory.
This theory has a mirror, which can be realized using the S-dual brane
configuration. 
As in examples with more supersymmetry, the Higgs branch of
the original theory gets mapped into the Coulomb branch of the mirror. 
We use parity invariance to argue that these branches cannot be lifted by
quantum corrections.
\end{abstract}

\newpage

\section{Introduction}

Mirror symmetry provides a relation between the infrared descriptions of
different supersymmetric gauge theories in three dimensions. The
first example of this kind was discovered in Ref.~\cite{is}. ${\cal N}=4$
supersymmetric QED with $N_f$ flavors has a dual description in terms of
another three dimensional gauge theory. The Coulomb branch of one theory is
mapped into the Higgs branch of the other and vice versa, which implies that 
the IR descriptions of these theories agree after suitable identifications. 
The exact change of variables relating these theories in the infrared is not
known, but recently some progress in that direction has been made
\cite{anton}.

While mirror symmetry was first discovered in a purely field theoretic 
context, it proved very useful to rederive and generalize the results of 
Ref.~\cite{is} using brane configurations in type IIB string theory \cite{hw} 
(see also \cite{dboer,pz} for different approaches and \cite{ohta} for a list of
such brane configurations). The three-dimensional
field theory is realized on D3-branes that terminate on NS5-branes and 
intersect D5-branes. The S-dual of this configuration gives rise to the 
mirror theory. 

Mirror symmetry can be generalized to include theories with ${\cal N}=2$
supersymmetry in three dimensions \cite{bhoy,bho,ah}. These theories can
also be realized on D3-branes in the presence of D5- and NS5-branes, and 
S-duality relates brane constructions that give rise to theories
that are mirror pairs. A different kind of IR duality between such
${\cal N}=2$ theories was discussed in \cite{ah0,karch}.

${\cal N}=4,2$ theories in three dimensions
can be obtained by dimensional reduction of four-dimensional ${\cal N}=2,1$
theories. Just like in the four-dimensional case, there are strong
constraints from R-symmetries, holomorphy, and non-renormalization theorems that
simplify the analysis of the vacuum structure. For theories with eight
supercharges (${\cal N}=4$ in three dimensions) it is possible to show that
the exact metric on the Coulomb branch of the original theory agrees with the
metric on the Higgs branch of the mirror. In the ${\cal N}=2$ case (four
supercharges) one can show that the complex structure of corresponding 
branches agrees.

In three dimensions one can also study ${\cal N}=1$ theories with two real
supercharges. These theories cannot be obtained by dimensional reduction of
higher dimensional supersymmetric field theories. The ${\cal N}=1$
superconformal algebra does
not contain any R-symmetries, there are no constraints from holomorphy,
and there are no know non-renormalization theorems. Nonetheless one can 
learn something about the infrared behavior of some of these theories
(see e.g.~\cite{aw,witten,bks}).

In this note we analyze three-dimensional ${\cal N}=1$ QED with $N_f$ flavors
and its mirror using 3-branes and 5-branes of type IIB string theory. This
theory does not have a Coulomb branch, since the ${\cal N}=1$ vector multiplet
does not contain any scalars. We argue that parity invariance prevents the
Higgs branch from being lifted by quantum corrections\footnote{This argument
is due to M.~Berkooz, A.~Kapustin, and M.~Strassler \cite{bks}.  }.
The S-dual brane configuration gives rise to the mirror theory
with a Coulomb branch that also cannot be lifted by quantum corrections. We 
suggest that these two theories are equivalent in the infrared. Unfortunately
it is hard to assemble additional evidence for this duality, since 
the metric on Higgs branch receives perturbative corrections at all orders
and it does not have a complex structure we could compare to that of the mirror
branch.
However, it seems highly plausible that a pair of S-dual brane configurations
should give rise to the same infrared physics on the 3-branes.

\section{${\cal N}=1$ Mirror Symmetry}

\begin{figure} 
\centerline{\epsfxsize = 8truecm \epsfbox{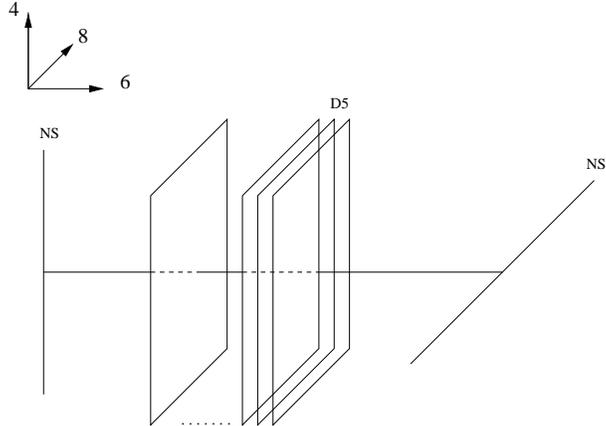} } 
\tighten{ 
\caption{ IIB brane configuration for ${\cal N}=1$ QED. The horizontal line
is the D3-brane and the other branes are labeled in the figure.
} 
\label{fig1} 
} 
\end{figure}

A brane configuration consisting of NS-branes in (012345), NS$'$-branes in
(012389), $N_f$ D5-branes in (012348), and D3-branes in (0126) preserves two 
supercharges and gives rise to a ${\cal N}=1$ theory in three dimensions.
The brane configuration is shown in Fig.~\ref{fig1}. It is straightforward
to determine the matter content of this theory. Each D5-brane provides one
${\cal N}=4$ hypermultiplet consisting of two complex scalars, $Q$,
$\tilde{Q}$, and fermions. We can combine these fields into a complex $SU(2)$
doublet ${\cal Q} = (Q,\tilde{Q}^\dagger)$. This $SU(2)$ is a remnant of the
$SU(2)_R$ symmetry of the ${\cal N}=4$ theory, but here it acts as an
ordinary global symmetry of the theory.
Without the D5-brane, this configuration is ${\cal N}=2$ supersymmetric,
so the theory contains an ${\cal N}=2$ vector multiplet which contains an
${\cal N}=1$ vector superfield and a real scalar superfield. The expectation
value of
the real scalar, $\phi$, corresponds to the $X_3$ position of the three brane.
In an
${\cal N}=2$ theory the scalar in the gauge multiplet couples to the matter
fields via terms of the form $|\phi Q|^2+|\phi \tilde{Q}|^2$
in the potential. 
In our configuration moving the 3-branes in the $X_3$ direction does not give 
a mass to the fundamentals, so our field theory differs from the ${\cal N}=2$
case by turning off this coupling.
Since we are considering a $U(1)$ theory,
this scalar is uncharged under the gauge group and we expect it to decouple from
the low energy dynamics of the gauge theory.

The brane configuration has no rotational symmetries, reflecting the fact that
the ${\cal N}=1$ superconformal algebra in three dimensions does not include an
R-symmetry. The Higgs branch of the theory can be seen geometrically by breaking
the 3-branes on the D5-branes and separating the pieces. For example, if there
is only one D5-brane, we can separate the two halfs of the 3-brane in the
$X_{3,4,8}$
directions. The $X_3$ position of the center of mass corresponds to the 
expectation value of the decoupled scalar. We expect a three-dimensional Higgs
branch in this case, because it can be parametrized by expectation value of the
fundamentals modulo gauge transformations, which eliminate one
of the four real parameters in ${\cal Q}$.

The theory has three real mass parameters for each flavor. In the brane picture
they correspond to motions of the D5-branes in $X_{5,7,9}$. These mass terms
appear in the Lagrangian in the same way as in a ${\cal N}=2$ theory, i.e., they
provide a complex and a real mass.
Unlike in the ${\cal N}=2$ case, the real mass cannot be
absorbed into a shift of the real scalar in the gauge multiplet, because there
is no coupling of this field to the fundamentals in the theory we discuss here.

This exhausts the list of possible brane motions. In particular there is no 
brane motion that corresponds to turning on FI parameters in the field theory.
This is consistent with our expectations, because the ${\cal N}=1$ vector 
multiplet does not contain any scalars.

To summarize, the classical 3-brane theory is a ${\cal N}=1$ supersymmetric
$U(1)$ gauge theory with $N_f$ complex fundamentals,
${Q}_i, \tilde{Q}^i$, a decoupled real scalar, $\phi$,
and no potential. The classical Higgs branch has $4N_f-1$ real dimensions. 
There are no non-renormalization theorems that protect the Higgs branch in 
theories with two supercharges, so the classical Higgs branch could be lifted 
by quantum corrections. However, discrete symmetries of the field theory 
provide some constraints \cite{aw,bks}.

Using the notation in Ref.~\cite{aw}, we can define a parity transformation,
$P$, that acts on the coordinates as $P(X_0,X_1,X_2)\to (X_0,-X_1,X_2)$
and on the spinors as $P\chi\to \pm \sigma_3\chi$. The Lagrangian of the
theory we described above is invariant under this transformation if all fields
are parity even. There is no parity anomaly, since the number of charged fields
is even. This ensures that parity is preserved in the full quantum theory.
The superpotential in a parity invariant theory must be parity
odd, because the the measure $d\theta^2$ is odd \cite{aw}.
Since the theory we consider here
does not contain any parity odd fields, we conclude that no superpotential 
can be generated.
This in turn implies that the Higgs branch cannot be lifted
by quantum corrections. Of course this does not mean that there are no quantum
corrections to the metric on the Higgs branch.

In order to find a candidate mirror theory of three-dimensional QED with $N_f$
flavors, we perform an S-duality on the brane configuration above. This turns
the NS-branes into D5-branes and vice versa. The field theory on the 3-brane
is now a $U(1)^{N_f-1}$ gauge theory. Each $U(1)$ factor contains an
${\cal N}=4$ vector multiplet, which decomposes into three real scalar 
superfields and one vector superfield in ${\cal N}=1$ language. There is a
${\cal N}=4$ hypermultiplet in the bifundamental representation at every
intersection of the 3-brane with an NS brane. These fields couple in the
${\cal N}=4$ supersymmetric way to the vector multiplets except at the first
and last NS brane. We discuss the $N_f=1,2$ cases is some detail. The
generalization to higher $N_f$ is straightforward. 

For $N_f=2$ the mirror is a $U(1)$ gauge theory with two ${\cal N}=4$ 
hypermultiplets $q_L$, $\tilde{q}_L$ and $q_R$, $\tilde{q}_R$ from strings
stretched across the left and right NS-branes respectively. The theory also
contains two uncharged complex scalars, $M_{L,R}$.
We identify their expectation values with the position of
the left and right sections of the 3-brane in $X_3+iX_4$ and $X_3+iX_8$
relative to the middle section.

Expectation values for the three scalars in the ${\cal N}=4$ gauge multiplet
correspond the the position of the middle section of the 3-brane in $X_{3,4,8}$.
Giving an expectation value to either the neutral scalar fields or the scalars
in the gauge multiplet makes the hypermultiplets massive. This implies a
superpotential of the form
\begin{eqnarray}
W &=& (M_L+\phi_3+i\phi_4)q_L\tilde{q}_L +
\phi_8(q_L^\dagger q_L-\tilde{q}_L\tilde{q}_L^\dagger) \\ \nonumber
&&+ (M_R+\phi_3+i\phi_8)q_R\tilde{q}_R +
\phi_4(q_R^\dagger q_R-\tilde{q}_R\tilde{q}_R^\dagger),
\end{eqnarray}
where the subscripts on the $\phi_i$ denote the directions they correspond to
in the brane construction. $\phi_3$ can be absorbed into a redefinition of
$M_{L,R}$, so it is again a free decoupled field. The Coulomb branch of this
theory can be parametrized by $M_{L,R}$, $\phi_{4,8}$ and the dual photon,
which provides seven real parameters in agreement with the count in the 
original theory. To ensure parity invariance of this theory, we take
the fundamentals to be parity even and the other fields to be parity odd.

The superpotential can be recast in a manifestly $SU(2)$ invariant way. The 
hypermultiplets combine into $SU(2)$ doublets
${\cal P}_{L,R} = (q_{L,R},\tilde{q}_{L,R}^\dagger)$, and the components of
$M_{L,R}$ and $\phi_{4,8}$ can be rearranged into two triplets $\Phi_{L,R}^i$.
In terms of these fields the superpotential reads
\begin{equation}
W = \Phi^i_L {\cal P}_L^\dagger \sigma_i {\cal P}_L +
\Phi^i_R {\cal P}_R^\dagger \sigma_i {\cal P}_R.
\end{equation}
The Coulomb branch can be parametrized by the expectation values of
$\Phi_{L,R}$. Since the fundamentals are massive on this branch, any 
dynamically generated superpotential can be expressed in terms of $\Phi_{L,R}$.
However, there are no parity odd and $SU(2)$ invariant combinations of these
fields. Thus we conclude that the Coulomb branch cannot be lifted by quantum
corrections and corresponds to the Higgs branch of the original theory. 

Moving the NS branes in $X_{5,7,9}$ corresponds to adding terms of the form
\begin{equation}
W_\xi = \Phi_L \cdot \xi_L + \Phi_R \cdot \xi_R
\end{equation}
to the superpotential. The $\xi_{L,R}$ are similar to FI parameters in the 
${\cal N}=4$ theory, because they force the matter fields to 
develope an expectation value. However, since the ${\cal N}=1$ vector
multiplet does not contain any scalars, they are not FI parameters in the 
original sense. Nonetheless we will abuse notation and refer to the fields
$\xi_{L,R}$ as FI parameters.

The FI parameters correspond to the three real mass terms in the original 
theory.
As a consistency check on our superpotential 
we turn on $\xi_L$ while keeping $\xi_R=0$. This
breaks the $U(1)$ gauge symmetry  and the resulting theory is a WZ
model with a real triplet $\Phi_R$, a complex doublet ${\cal P}_R$, and a 
superpotential
\begin{equation}
W = \Phi^i_R {\cal P}_R^\dagger \sigma_i {\cal P}_R.
\end{equation}
This agrees with the superpotential we read off 
from the S-dual of the $N_f=1$ brane configuration,
\begin{equation}
W = M_L q \tilde{q} + {\rm Im}[M_R] (qq^\dagger - \tilde{q}^\dagger\tilde{q}),
\end{equation}
up to field redefinitions. 

We can readily construct non-Abelian theories using our brane configuration. If
we put $N_c$ 3-branes between the NS-branes, we get a $SU(N_c)$ gauge theory
in three dimensions. However, the analysis of the field theory in this case
is much more involved. The scalar in the ${\cal N}=2$ vector multiplet no longer
decouples, since it is now charged under the gauge group. Since it 
does not couple to the fundamentals, they do not become massive if we give
an expectation value to this field. This implies that we can move onto the
Higgs branch (parametrized by the expectation value of the fundamentals) 
at any point on the Coulomb branch (parametrized by the expectation value
of the adjoint scalar). Thus classically we find an interacting conformal
field theory
at every point of the moduli space, which is very difficult to analyze.
We will not discuss the non-Abelian case any further in this paper.

\section{Discussion}

In this note we discussed a type IIB brane configuration that gives rise to 
${\cal N}=1$ QED with $N_f$ flavors in three dimensions. We argued that the
Higgs branch of this theory cannot be lifted by quantum corrections, using the
parity invariance of the theory. Since there are no non-renormalization 
theorems for theories with two supercharges, the metric on the Higgs branch
receives quantum corrections but our argument indicates that this does not
change its dimension. 

For ${\cal N}=2,4$ theories, S-dual pairs of brane configurations give rise
to three dimensional theories that are mirror pairs. We suggest that this
is true also for the ${\cal N}=1$ theory we consider here. To support this
claim we showed that the S-dual brane configuration gives rise to another 
${\cal N}=1$ theory that has a Coulomb branch of the same dimension as the 
Higgs branch of
original theory. Parity invariance ensures that neither of these branches 
can be lifted by quantum corrections. The mass parameters of the original
theory are mapped into FI parameters of the mirror. 
To the best of our knowledge, this is the first proposed duality between
theories with two supercharges.

Unfortunately the low degree of supersymmetry makes it hard to
check our proposal any further. The metric on the 
moduli spaces on both sides have quantum corrections and to establish their
equivalence, these would have to be computed. Since these branches are real
manifolds, there are no complex structures we can compare. 
On the other hand, some support for our claim comes from the fact that the 
version of three dimensional QED we discuss here is very similar to the
theory one obtains by giving a mass to all scalars in the 
vector multiplet of ${\cal N}=4$ QED. The only difference between this theory 
and the theory we discussed here is the presence of the decoupled scalar in 
our case. Since it does not participate in the dynamics we expect these two
theories to be closely related.
In fact, adding a term $W = m^i \phi_i$, where $m^i$ are uncharged scalars and
$\phi_i$ are the three scalars in the ${\cal N}=4$ vector multiplet to the
superpotential of ${\cal N}=4$ QED and perturbing the mirror by the
corresponding FI terms, generates the mirror pairs we discussed above
\cite{bks}.
While this lends some support to our claims, it is
not a very strong argument. Theories with two supercharges can undergo phase
transitions as we vary the mass of the three scalars in the ${\cal N}=4$
vector multiplet, so it is not clear that perturbing
a pair of mirror theories with ${\cal N}=4$ supersymmetry causes them to flow
to mirror pairs of ${\cal N}=1$ theories. The analysis in this paper suggests
that this does happen in the specific case we study here.

\acknowledgments

It is a pleasure to thank Anton Kapustin for several very helpful discussions.
This work was supported in part by DOE grants \#DF-FC02-94ER40818 and
\#DE-FC-02-91ER40671.

{\tighten

}

\end{document}